# Crossover from antiferromagnetic to ferromagnetic ordering in semi-Heusler alloys $Cu_{1-x}Ni_xMnSb$ with increasing Ni concentration

Madhumita Halder, S. M. Yusuf,* and Amit Kumar

*Solid State Physics Division, Bhabha Atomic Research Centre, Mumbai 400085, India*

A. K. Nigam

*Tata Institute of Fundamental Research, Homi Bhabha Road, Mumbai 400005, India*

L. Keller

*Laboratory for Neutron Scattering, Paul Scherrer Institut, CH-5232 Villigen PSI, Switzerland*

**Abstract**

The magnetic properties and transition from an antiferromagnetic (AFM) to a ferromagnetic (FM) state in semi Heusler alloys $Cu_{1-x}Ni_xMnSb$, with $x < 0.3$ have been investigated in details by dc magnetization, neutron diffraction, and neutron depolarization. We observe that for $x < 0.05$, the system $Cu_{1-x}Ni_xMnSb$ is mainly in the AFM state. In the region $0.05 \leq x \leq 0.2$, with decrease in temperature, there is a transition from a paramagnetic to a FM state and below ~50 K both AFM and FM phases coexist. With an increase in Ni substitution, the FM phase grows at the expense of the AFM phase and for $x > 0.2$, the system fully transforms to the FM phase. Based on the results obtained, we have performed a quantitative analysis of both magnetic phases and propose a magnetic phase diagram for the $Cu_{1-x}Ni_xMnSb$ series in the region $x < 0.3$. Our study gives a microscopic understanding of the observed crossover from the AFM to FM ordering in the studied semi Heusler alloys $Cu_{1-x}Ni_xMnSb$.

* Electronic mail: smyusuf@barc.gov.in FAX: +91 22 25505151

# I. INTRODUCTION

Heusler and semi-Heusler alloys have become a subject of investigation, both theoretically and experimentally, in recent years because of their interesting physical properties.[1-4] This class of materials has become potential candidates for spintronics application because of their half metallic character, structural similarity with semiconductors, and Curie temperature above room temperature. Semi-Heusler alloy NiMnSb is one of the best known examples of half metallic ferrromagnets. de Groot *et al.*,[5] based on electronic structure calculation, predicted that NiMnSb should exhibit 100% spin polarization at the Fermi level. Other interesting physical property of this class of materials is the martensitic transformation at low temperatures,[6] which gives rise to some interesting properties like magnetic shape memory effect, inverse magnetocaloric effect, *etc.*[1, 4] These properties are promising for future technological applications. From basic understanding point of view, these systems show a rich variety of magnetic behaviors ranging from itinerant to localized magnetism with a wide diversity in the magnetic properties like ferromagnetism, ferrimagnetism, antiferromagnetism and other types of noncollinear ordering.[7-12]

The semi-Heusler alloys *X*MnSb ($X = 3d$ elements) belong to a class of materials with high local magnetic moments on the Mn atoms. The Mn-Mn distance in these alloys is fairly large ($d_{\text{Mn-Mn}} > 4$ Å) for a direct exchange interaction to propagate. The magnetic exchange interaction in the Mn-based semi-Heusler alloys varies from ferromagnetic (FM) Ruderman-Kittel-Kasuya-Yosida (RKKY) type exchange to antiferromagnetic (AFM) superexchange interactions with Sb (*sp*) and *X* (3*d*) atoms playing a role in mediating the exchange interactions between Mn atoms.[12] The semi-Heusler alloy NiMnSb is a ferromagnet with Curie temperature of $T_C = 750$ K.[13] It crystallizes in the C1$_b$ structure with four interpenetrating fcc sub-lattices.[13] The band structure calculations show that the magnetic properties of NiMnSb are due to the magnetic moments localized only on the Mn atoms interacting via itinerant electrons in the conduction band i.e. the exchange mechanism is of RKKY type.[14] CuMnSb alloy also has the same crystal structure but antiferromagnetic with Néel temperature $T_N = 55$ K.[15] The magnetic moment is only on the Mn atom and is aligned perpendicular to the ferromagnetic (111) planes with neighboring planes orientated in antiparallel.[16] In case of CuMnSb, the exchange interaction is of superexchange type. The AFM to FM phase transition in $Cu_{1-x}Ni_xMnSb$ is a consequence of the dominance of ferromagnetic RKKY-type exchange interaction over the antiferromagnetic superexchange interaction which occurs by tuning of the $X$ (nonmagnetic 3$d$ atoms Cu/Ni). Change in the electron concentration (i.e. difference in Cu and Ni valencies), modifies the density of states at the Fermi surface, which affects the exchange interaction between Mn-Mn spins in the Mn sublattice, resulting in the AFM to FM transition. There are reports on electronic, magnetic, and transport properties and on the magnetic phase transition in $Cu_{1-x}Ni_xMnSb$, both theoretically[17, 18] and experimentally[15, 19, 20] which show that there is a decrease in magnetization and electrical conductivity with decreasing $x$, for $x < 0.3$. For $x > 0.3$, compounds of the $Cu_{1-x}Ni_xMnSb$ series are ferromagnetic in nature with a nearly constant value of Mn moment (~ 4 $\mu_B$/atom). The magnetic ordering temperature increases continuously with $x$ for the entire series. The theoretical studies, based on the density-functional theory, have shown that for $x < 0.3$, antiferromagnetic superexchange coupling dominates [17]and the FM phase decays into a complex magnetic phase which can be viewed as the onset of disorder in the orientation of the Mn spins.[18] However, there is no detailed experimental study reported in the $x < 0.3$ region of $Cu_{1-x}Ni_xMnSb$ series, where the transition from the AFM to the FM state occurs. Moreover, the reported experimental studies[15, 19, 20] are based on bulk techniques such as magnetization and resistivity. There is no microscopic understanding of the nature of AFM to FM phase transition. This motivated us to investigate the $Cu_{1-x}Ni_xMnSb$ series in detail, in the region $x < 0.3$, by dc magnetization, neutron diffraction and neutron depolarization techniques in order to have a detailed understanding of the nature of this AFM to FM transition. Our results suggest electronic phase separation in the $0.05 \leq x \leq 0.2$ region i.e. both AFM and FM phases coexist. Our study also gives a quantitative analysis for both magnetic phases and a magnetic phase diagram for the $Cu_{1-x}Ni_xMnSb$ series in the $x < 0.3$ region. The present study will be useful for understanding the nature of magnetic ordering as well as for tuning magnetic and electronic properties of different Heusler and semi-Heusler alloys.

## II. EXPERIMENTAL DETAILS

The polycrystalline $Cu_{1-x}Ni_xMnSb$ samples ($x$ = 0.03, 0.05, 0.07, 0.15 and 0.2) with constituent elements of 99.99 % purity, were prepared by an arc-melting under argon atmosphere. An excess of Mn and Sb (2 wt. %) was added to the starting compositions to compensate the evaporation losses. For better chemical homogeneity, the samples were re-melted many times. After melting, they were annealed in vacuum-sealed quartz tube at 650 ºC for 7 days. The powder x-ray diffraction (XRD) using the Cu-$K_\alpha$ radiation in the 2θ range of 10 – 90º with a step of 0.02º was carried out on all samples at room temperature. The dc magnetization measurements were carried out on the samples using a SQUID magnetometer (Quantum Design, MPMS model) as a function of temperature and magnetic field. The zero-field-cooled (ZFC) and field-cooled (FC) magnetization measurements were carried out over the temperature range of 5–300 K under 200 Oe field. Magnetization as a function of magnetic field was measured for $x$ = 0.05, 0.07, 0.15, and 0.2 samples at 5 K over a field variation of ± 50 kOe. Neutron diffraction patterns were recorded at various temperatures over 5–300 K for $x$ = 0.03, 0.05, 0.07, 0.15, and 0.2 samples using the powder diffractometer – II (λ = 1.2443 Å) at the Dhruva reactor, Trombay, Mumbai, India. For the $x$ = 0.15 sample, the temperature dependent neutron diffraction experiments were also performed down to 1.5 K on the neutron powder diffractometer DMC (λ = 2.4585 Å) at the Paul Scherrer Institute (PSI), Switzerland. The one-dimensional neutron-depolarization measurements were carried out for $x$ = 0.03, 0.05, 0.07, and 0.15 samples down to 2 K using the polarized neutron spectrometer (PNS) at the Dhruva reactor (λ = 1.205 Å). FC neutron-depolarization measurements were carried out by first cooling the sample from room temperature down to 2 K in the presence of 50 Oe field (required to maintain the neutron beam polarization at sample position) and then carrying out the measurements in warming cycle under the same field. The incident neutron beam was polarized along the –z direction (vertically down) with a beam polarization of 98.60(1)%. The transmitted neutron beam polarization was measured along the +z direction as described in an earlier paper.[21]

## III. RESULTS AND DISCUSSION

Figure 1(a) shows the XRD patterns for all the samples at room temperature. The Rietveld refinement (using the FULLPROF program [22]) of the XRD patterns at room temperature confirms that all samples are in single phase with $C1_b$ type cubic structure and space group $F\bar{4}3m$. From the Rietveld refinement, we find that Cu/Ni atoms occupy the sub-lattice (000), while Mn and Sb atoms occupy other two sub-lattices (¼ ¼ ¼) and (¾ ¾ ¾), respectively, as known for the $C1_b$ type cubic structure.[16] The fourth sublattice (½ ½ ½) is unoccupied. From the Rietveld refinement, we confirm that (½ ½ ½) sublattice is unoccupied for all samples. Since Mn, Ni and Cu are nearby elements in the periodic table, XRD technique is not sensitive enough to confirm any interchange of the Cu/Ni and the Mn site atoms. The absence of the interchange of the atoms between (000) and (¼ ¼ ¼) sites (generally present in these types of structure) has been confirmed by neutron diffraction study (discussed later). The present XRD study shows that the lattice parameter decreases with increasing Ni substitution in CuMnSb [Fig. 1(b)].

### A. dc magnetization study

Figure 2 shows the ZFC and FC magnetization (*M*) vs temperature (*T*) curves under an applied field of 200 Oe for the $x$ = 0.03, 0.05, 0.07, 0.15 and 0.2 samples. An antiferromagnetic peak is observed at around 54, 51, and 50K for the $x$ = 0.03, 0.05, and 0.07 samples, respectively, in both FC and ZFC $M(T)$ curves which can be estimated as the antiferromagnetic transition temperature. However, for the $x$ = 0.05 and 0.07 samples, a bifurcation (in the FC and ZFC curves) is observed below ~ 45 K. For the FC case, the magnetization attains a constant value at lower temperatures indicating the presence of some ferromagnetic contribution for both $x$ = 0.05 and 0.07 samples. For higher Ni substitution (the $x$ = 0.15 sample), the antiferromagnetic peak is still present, however, it becomes broad. The FC and ZFC curves show a bifurcation only below ~ 45 K. A bifurcation in FC and ZFC curves and a constant value of magnetization in FC $M(T)$ curve are expected in these compounds when competing AFM and FM interactions are present. The constant value of magnetization in FC curves of the $x$ = 0.05, 0.07, and 0.15 samples below 45 K indicates that on substituting Ni in the AFM CuMnSb, some FM like clusters appear in the AFM matrix. During the ZFC process, the FM clusters freeze in random directions resulting in the random orientation of magnetization of individual clusters. While in the FC process, the FM clusters align along the direction of the applied field and contribute to a higher and constant value of magnetization below 45 K. On further increase in the Ni concentration i.e. for the $x$ = 0.2 sample, the antiferromagnetic peak disappears and a negligible bifurcation between FC and ZFC curves occurs, indicating that the nature of the $M(T)$ curve is that of a typical ferromagnetic system. Also the value of magnetization increases with increasing Ni concentration. The transition temperature also increases with Ni substitution as reported in litrature.[19]

Figure 3(a) shows the $M$ vs applied field ($H$) curves at 5 K over a field range of ± 50 kOe (all four quadrants) for $x$ = 0.05, 0.07, 0.15 and 0.2 samples. The enlarged view of low field region of $M$ vs $H$ curves for $x$ = 0.05 and 0.07 samples, is shown in the top left inset of Fig. 3(a) and for $x$ = 0.15 and 0.2 samples, it is shown in the bottom right inset of Fig. 3(a). The observed hysteresis for all four samples indicates the presence of ferromagnetism. The value of magnetization ($M_{Max}$), at the maximum applied field (50 kOe) in our study, increases with increase in Ni concentration [shown in Fig. 3(b)], indicating that FM phase increases with increase in Ni concentration. The coercive field [shown in Fig. 3(b)] first increases as we increase Ni substitution from $x$ = 0.05 to 0.07 which could be due to large anisotropy of the isolated FM like clusters in these samples and then decreases with further increase in Ni concentration ($x$ = 0.07 to 0.2) indicating a soft ferromagnetic nature for high Ni concentration samples. The Arrott plots for the $x$ = 0.05 and 0.07 samples at 5 K are shown in Fig. 3(c). The linear extrapolation of the Arrott plots at high fields, intercepts the negative $M^2$ axis, indicating that there is no spontaneous magnetization for these samples. Whereas the presence of spontaneous magnetization is evident from the Arrott plots for the $x$ = 0.15 and 0.2 samples (shown in Fig.3 (d)), indicating a dominant FM nature of these samples. The volume fractions of the AFM and FM states have been estimated for the $x$ = 0.05, 0.07, 0.15 and 0.2 samples from their maximum magnetization, ($M_{Max}$) values at 50 kOe. From band structure calculation it was concluded that in FM NiMnSb, the moment was localized on Mn and the net moment was found to be 3.96 $\mu_B$/f.u.[2], 4 $\mu_B$/f.u [23] and 4 $\mu_B$/f.u. [24] This is due to the large exchange splitting of the Mn 3$d$ electrons in comparison to the Ni 3$d$ electrons which are weekly spin polarized. Experimentally found moment per Mn atom for NiMnSb is 3.85 $\mu_B$/atom.[25] In case of AFM CuMnSb as well, only Mn carries the moment and is ~ 3.9 $\mu_B$/atom.[16] In the present study, $M_{Max}$ for the $x$ = 0.2 sample is 3.5 $\mu_B$/f.u, whereas the expected moment is ~3.85 $\mu_B$/f.u. for 100% FM $Cu_{1-x}Ni_xMnSb$ samples considering that only Mn atoms carries the moment. So the volume fraction of FM phase for the $x$ = 0.2 sample is about 91%. The remaining fraction can be considered as the AFM phase. Similarly FM and AFM phase fractions have been calculated for the $x$ = 0.05, 0.07, and 0.15 samples and FM phase fractions are around 10, 17, and 60 %, respectively.

## B. Neutron diffraction study

To study the transition from AFM to FM state in $Cu_{1-x}Ni_xMnSb$ semi-Heusler alloys, in details, we performed the neutron diffraction study on the $x$ = 0.03, 0.05, 0.07, 0.15, and 0.2 samples at various temperatures in magnetically ordered as well as paramagnetic (PM) states. The measured diffraction patterns were analyzed by the Rietveld refinement technique using the FULLPROF program.[22] The reported values of the atomic positions and lattice constants for CuMnSb were used as the starting values for the present Rietveld refinement for all samples.[16] The analysis reveals that the crystal structure has four interpenetrating fcc sub-lattices i.e. $C1_b$ type structure as observed in the XRD. Here neutron diffraction easily distinguishes between Ni and Cu due to difference in their scattering lengths ($1.03\times 10^{-12}$ and $0.77\times10^{-12}$ cm for Ni and Cu, respectively). We confirm that entire Ni is substituted at the Cu site. The low temperature (at 5 K) neutron diffraction patterns for the $x$ = 0.03, 0.05 and 0.07 samples (Fig. 4), show a number of additional Bragg peaks when compared with diffraction patterns recorded in PM state for these samples. These peaks appear below 50 K and can be indexed in terms of an antiferromagnetic unit cell having lattice parameters twice that of the chemical unit cell, similar to that of CuMnSb, with magnetic moments aligned perpendicular to the ferromagnetic (111) planes and neighboring planes oriented in antiparallel.[16] The $x$ = 0.03 sample shows a pure AFM phase with a moment of 3.14(3) $\mu_B$ per Mn [Table 1]. However, for $x$ = 0.05 and 0.07 samples, a small ferromagnetic phase contribution (~10% and ~17 %, respectively) was obtained from dc magnetization data, which could not be detected in the neutron diffraction data possibly due to the low neutron flux at our instrument. Therefore, for the magnetic refinement, only AFM phase has been considered with 100% phase fraction (for $x$ = 0.05 and 0.07) and the derived values of the Mn moment per atom are given in Table 1. The diffraction pattern for the $x$ = 0.2 sample [Fig. 4(g)] shows no extra peaks at low temperature (5 K) but an extra Bragg intensity to the lower angle fundamental (nuclear) Bragg peaks has been observed, apparently indicating a pure ferromagnetic nature of the sample. However from our dc magnetization study for the $x$ = 0.2 sample it was concluded that ~ 9% volume fraction of the sample is AFM. This small AFM phase fraction could not be detected in our neutron diffraction study. Therefore, only FM phase has been considered (100 %) to derive Mn moment per atom [Table 1]. For the $x$ = 0.15 sample, neutron diffraction measurements carried out at Paul Scherrer Institute (PSI), Switzerland at 1.5 K [Fig. 5(a)], show a number of additional Bragg peaks ( as compared to the diffraction patterns at 50 K and above) as well as observable extra Bragg intensity to the lower angle fundamental (nuclear) peaks. The extra Bragg peaks can be indexed to an antiferromagnetic structure similar to that found for other samples ($x$ = 0.03, 0.05, and 0.07). These extra peaks disappear above 50 K. The difference pattern obtained by subtracting 250 K data (PM state) from 1.5 K data is shown in Fig. 5(d). Both AFM and FM contributions to the intensity are observed. The difference pattern obtained by subtracting 250 K data from 50 K data [Fig. 5(e)] shows only the FM contribution to the intensity at 50 K. In this case, for the magnetic refinement at 1.5 K, we have considered both FM and AFM phases, and the corresponding Mn moment per atom for each phase is given in Table 1. The magnetic phase fraction has also been derived. The ferromagnetic moment for $x$ = 0.15 and 0.2 samples is found to align along the crystallographic-axes. The corrected FM Mn moment for the $x$ = 0.2 sample, derived by considering the appropriate phase fraction, is 3.66(5) $\mu_B$. The corrected values of AFM Mn moment, obtained by considering the appropriate AFM phase fraction estimated from magnetization data, are 3.12(4) and 3.05(7) $\mu_B$ at 5 K for $x$ = 0.05 and 0.07 samples, respectively. Here, for estimation of the Mn moment, we have considered the fact that for a given intensity in neutron diffraction, the moment is inversely proportional to the square root of the scale factor (volume phase fraction) in the Rietveld refinement.[22] The lesser value of the AFM Mn moment at 5 K could be to due to the large value of $T/T_N$ ratio. The neutron diffraction study, therefore, indicates that there is a crossover from an AFM to a FM state on substituting Ni in CuMnSb. It is also evident that the derived values of both AFM and FM Mn moments (after correcting for appropriate phase fractions as obtained from dc magnetization data) remain almost constant across the studied series. The appearance of FM Mn moment in AFM CuMnSb on substituting Ni may be viewed as some of the AFM Mn spins change their direction and align parallel to each other. These uncompensated spins align (parallel) along the crystallographic axes and can be treated as FM like clusters in the AFM matrix. This is in agreement with the theoretical model of uncompensated disordered local-moment proposed by Kudrnovský *et al.*[18] As more and more Mn atoms align (parallel) with increasing Ni concentration, the disordered AFM moment appears in equal amount as FM moment. There could be two possible models for the transition from the AFM to FM state. First is the inhomogeneous model, where in the intermediate ($0.05 \leq x \leq 0.2$) concentration region, both AFM and FM phases coexist.[26] As we change the Ni concentration from $x$ = 0.05 to 0.2, the volume fraction of the two phases changes. The second model is the homogeneous model, where the AFM and FM contributions to the Mn moment result from a canted magnetic structure.[26] Szytula proposed a canted spin structure for the AFM to FM phase transition in $Cu_{1-x}Ni_xMnSb$.[27] It was suggested that the magnetic moments are canted at an angle of 45° to the cube edge, which when resolved into components would give both AFM and FM contributions. We have tried to analyze the neutron diffraction data with both models and find that the inhomogeneous model with phase coexistence only fits the data. A canted behavior was not observed from the analysis of the neutron diffraction data for any of the present samples as evident from nearly constant and almost full values of the derived FM and AFM Mn moments. If a canted structure exists as suggested by Szytula, then the vertical component of the moment (along the crystallographic-axes), as found in our neutron diffraction study, would give the FM contribution and the horizontal component (in the basal plane) should give the AFM contribution. Therefore the AFM and FM moments should be perpendicular to each other. But in the present case, the intensity of the antiferromagnetic peaks fits only if we consider the moment direction to be perpendicular to the (111) plane that makes an angle of 45° with FM moment. Furthermore, for a canted structure, the disappearance of the antiferromagnetic order at higher temperature (above 50 K) would indicate that the magnetic moment changes its direction and the system becomes a collinear ferromagnet. In that case, there would be a sudden increase in the intensity of the FM Bragg peaks. We have plotted the integrated intensities of the (111), (200) and (220) nuclear Bragg peaks for the $x = 0.15$ sample as a function of temperature [shown in Fig. 5(g)]. These nuclear Bragg peaks have finite contribution to the intensity arising from the FM ordering of Mn moments. We find that the intensity of these Bragg peaks gradually decreases with increasing temperature. So our neutron diffraction data infer the coexistence of both AFM and FM phases (for $0.05 \leq x \leq 0.2$) under the inhomogeneous model. Figure 6 shows the variation of the AFM and the FM phases with the Ni concentration at low temperature as obtained from dc magnetization and neutron diffraction experiments. Phase fraction obtained from the analysis of neutron diffraction data for $x = 0.15$ sample is close to that of the results obtained from magnetization study. A coexistence of two magnetic phases i.e. AFM and FM was reported in $Cu_{1-x}Pd_xMnSb$[28] and $Pd_2MnSn_xIn_{1-x}$[26] Heusler alloys series. But no quantitative analysis of magnetic phases and their evolutions as a function of temperature or increasing atomic substitution were studied. Our study gives a microscopic understanding of the AFM to FM phase transition and the variation of the two phases as a function of Ni concentration in the $Cu_{1-x}Ni_xMnSb$ series. Moreover, a magnetic phase diagram in the temperature and Ni-concentration plane is proposed here.

### C. Neutron Depolarization Study

The dc magnetization and neutron diffraction experiments indicate the appearance of FM clusters/domain in the AFM matrix with increasing Ni content in CuMnSb. To study such type of magnetic inhomogenities (FM clusters/domains in the AFM matrix) on a mesoscopic length scale, neutron depolarization is a powerful technique. We have carried out the one dimensional neutron depolarization study for the $x$ = 0.03, 0.05, 0.07, and 0.15 samples. Typically, the neutron depolarization results for various kinds of magnetic materials are as follows.[29-33] In case of an unsaturated ferromagnetic or ferrimagnetic material, the magnetic domains exert a dipolar field on the neutron polarization and depolarize the neutrons due to Larmor pression of the neutron spins in the magnetic field of domains. In pure antiferromagnetic materials, there is no net magnetization, hence no depolarization occurs. In a paramagnetic material, the neutron polarization is unable to follow the variation in the magnetic field as the temporal spin fluctuation is too fast ($10^{-12}$ s or faster). Hence, no depolarization is observed. However, depolarization is expected in case of clusters of spins with net magnetization.[21, 31-33] Thus, neutron depolarization technique is a mesoscopic probe which detects the magnetic inhomogenities in the length scale of 100 Å to several microns. Figure 7 shows the temperature dependence of the transmitted neutron beam polarization $P$ for an applied field of 50 Oe applied parallel to the incident neutron beam polarization. For $x$ = 0.03 sample, there is no change in value of $P$ (shown in the main as well as in the inset of Fig. 7), which indicates that the sample is antiferromagnetic in nature. For the $x$ = 0.05 sample, shown in the main as well as in inset of Fig. 7, there is slight decrease in the value of $P$ at $T < 70$ K indicating the presence of small ferromagnetic-like clusters in the antiferromagnetic matrix below ~ 70 K. For further increase in Ni concentration i.e. for $x$ = 0.07 and 0.15 samples, $P$ shows a continuous decrease from ~76 K and ~173 K, respectively. $P$ attains constant values below ~ 50 K indicating that there is no further growth of the domains at lower temperatures. The temperature at which the value of $P$ starts decreasing can be considered as the ferromagnetic transition temperature $T_C$. The neutron beam polarization in a depolarization experiment can be represented by the following expression[29, 34]

$$P = P_i \exp\left[-\alpha\left(\frac{d}{\delta}\right)\langle\Phi_\delta\rangle^2\right], \qquad (1)$$

where $P_i$ and $P$ are the initial and transmitted neutron beam polarization, α is a dimensionless parameter ≈ ⅓, $d$ is effective sample thickness, δ is the average domain size, and $\Phi_\delta = (4.63\times10^{-10}\ \mathrm{G^{-1}\ Å^{-2}})\lambda\delta B$ the precession angle. Here λ is the wavelength of the neutron beam and $B$ (= $4\pi M_S\rho$, $M_S$ being spontaneous magnetization and ρ, density of the material) is the average magnetic induction of a domain/cluster. The above equation is valid only when the precession angle $\Phi_\delta$ is a small fraction of 2π over a typical domain/cluster length. The increasing observed neutron beam depolarization at temperatures below $T_C$ with increasing the $x$ indicates the presence of larger ferromagnetic domains/clusters consistent with neutron diffraction and dc magnetization experiments. It may be noted here that, the Arrott plots for the $x$ = 0.05 and 0.07 compounds [Fig. 3(c)] do not yield any spontaneous magnetization. However we have observed neutron depolarization for these samples. This interesting observation indicates about the dynamics of the FM clusters in these compositions. Generally, neutron polarization vector senses fluctuating magnetic fields averaged over a time scale of the order of Larmor precession time which is of the order of $10^{-8}$ seconds for 1 kG magnetic induction ($B$) of the domain. So, if the fluctuation time of these FM clusters/domains is larger than the Larmor precession time, one would get neutron depolarization from such clusters/domains. dc magnetization measurements, on the other hand, are time averaged (over several seconds) measurements resulting in a zero spontaneous magnetization over the experimental time scale. Neutron depolarization study indicates that with increase in the Ni concentration in CuMnSb i.e. from $x$ = 0.05 to 0.07, the ferromagnetic like clusters with net magnetic moment (in the antiferromagnetic matrix) increase in size. Further increase in the Ni concentration, drives the system towards a FM state. This is consistent with the results obtained from dc magnetization and neutron diffraction experiments

The magnetic behavior in $Cu_{1-x}Ni_xMnSb$ semi-Heusler alloys can be interpreted in terms of a delicate balance between the two competitive exchange interactions i.e. ferromagnetic RKKY-type exchange and antiferromagnetic superexchange interaction. Here, we observe that the FM state appears in AFM CuMnSb with small substitution of Ni at the Cu site (i.e. $x = 0.05$). Our results suggest an electronic phase separation in the $0.05 \leq x \leq 0.2$ region. The quenched disorder in the Cu/Ni sublattice causes a disorder in the orientation of the spins at the Mn sublattice.[18] As we increase the Ni concentration, more and more spins align parallel i.e. the FM clusters grow in size and finally the system becomes completely ferromagnetic. This is evident from the dc magnetization, neutron diffraction, and neutron depolarization studies. Based on the results of our experimental studies, we propose a magnetic phase diagram of the present $Cu_{1-x}Ni_xMnSb$ ($x$ = 0 to 1) semi-Heusler alloys series (shown in Fig. 8). The values of Curie temperature for some of the samples of the series are taken from Ref. 19. The Néel temperature for the $x$ = 0.05, 0.07 and 0.15 samples and the Curie temperature for the $x$ = 0.2 sample have been obtained from our dc magnetization data, while Curie temperature for the $x$ = 0.07 and 0.15 samples have been obtained from our neutron depolarization data. We observe that only a narrow region of $x$ ($\leq 0.05$) has the pure AFM phase and in the region $0.05 \leq x \leq 0.2$, with decrease in temperature, there is a transition from PM to FM state and below ~50 K both AFM and FM phases coexist. With increase in $x$ further, most part of the phase diagram is dominated by the FM phase. Similar results were observed in $Pd_2MnSn_xIn_{1-x}$ by Khoi *et al*,[26] which is an example of 'bond randomness' transition in a Heisenberg system, where NMR data showed the coexistence of both AFM and FM domains. Their results showed that intermediate region had an inhomogeneous structure with two different types of coexisting order (AFM and FM) separated in space. The theoretical phase diagram of such quenched random alloys where one end is FM and other AFM, shows that the two phases are either separated by a mixed phase or by a first-order phase line.[35] Our results suggest that crossover from AFM to FM transition in $Cu_{1-x}Ni_xMnSb$ series is separated by a mixed phase. A possible reason could be due to the long-range nature of the exchange interaction (RKKY-type) present in the system. The transition from the AFM state to FM state occurs continuously with increase in the Ni content and no abrupt change is observed. The electronic structure calculation shows that, with increase in the Ni substitution, the ferromagnetic RKKY-type exchange increases due to an increase in spin polarization of the conduction electrons at the Fermi-level, and the superexchange interaction decreases as the Fermi energy moves away from the unoccupied Mn 3$d$ density of states.[18] As a result there is a crossover from AFM to FM state in the $Cu_{1-x}Ni_xMnSb$ with increase in Ni concentration.

## IV. SUMMARY AND CONCLUSIONS

Here, we have investigated the $Cu_{1-x}Ni_xMnSb$ series in the region $x < 0.3$, to bring out the microscopic nature of the AFM to FM transition, by dc magnetization, neutron diffraction and neutron depolarization techniques. We observe that the FM state appears in AFM CuMnSb with small substitution of Ni at the Cu site (i.e. 5%). We find that below $x$ = 0.05, the system is mainly in the AFM state. In the region $0.05 \leq x \leq 0.2$, with decrease in temperature, there is a transition from PM to FM state and below ~ 50 K both AFM and FM phases coexist. Above $x$ = 0.2, the system is mainly in the FM state. The FM state can be viewed as some of the antiferromagnetically aligned Mn spins change their orientation and align parallel to each other forming a FM like cluster in the AFM matrix. These clusters grow in size with increase in Ni content as more and more spins align parallel and finally drives the system to the FM state. The results are consistent with the reported electronic band structure calculation. The results of the present investigation show a path in tuning magnetic and electronic properties of different Heusler and semi-Heusler alloys for various practical applications and can be used to fabricate materials with desired physical properties.

## ACKNOWLEDGMENT

M. H. acknowledges the help provided by A. B. Shinde and A. Das for the neutron diffraction data collection and K. Shashikala, M. D. Mukadam, and A. K. Rajarajan for sample preparation.

**Table 1**

Derived FM and AFM moments per Mn atom from neutron diffraction data for various compositions

| $Cu_{1-x}Ni_xMnSb$ | Temperature (K) | FM Moment ($\mu_B$) | AFM Moment ($\mu_B$) |
|---|---|---|---|
| $x = 0.03$ | 5 | | 3.14(3) |
| $x = 0.05$ | 5 | -* | 2.74(4) |
| $x = 0.07$ | 5 | -* | 2.42(2) |
| $x = 0.15$ | 1.5 | 3.75(4) | 3.38(3) |
| $x = 0.2$ | 5 | 3.48(4) | -* |

* Magnetic moment per Mn atom for this phase is below the detection limit of our neutron diffraction experiment.

**List of Figures**

FIG. 1: (a) x-ray diffraction patterns for $x$ = 0.03, 0.05, 0.07, 0.15 and 0.2 samples at room-temperature. The (*hkl*) values corresponding to Bragg peaks are marked. (b) Variation of lattice constant with Ni concentration.

FIG. 2: (Color online) Temperature dependence of FC and ZFC magnetization $M$ for $x$ = 0.03, 0.05, 0.07, 0.15, and 0.2 samples at 200 Oe applied field.

FIG. 3: (Color online) (a) The $M$ vs $H$ curves over all the four quadrants for $x$ = 0.05, 0.07, 0.15, and 0.2 samples at 5 K. Inset (i) shows the enlarge view of the $M$ vs $H$ curves for $x$ = 0.05 and 0.07 samples, where a clear hysteresis is observed. Inset (ii) shows the enlarge view of the $M$ vs $H$ curves for $x$ = 0.15 and 0.2 samples. (b) The variation of maximum magnetization (at 50 kOe) and coercive field with Ni concentration. Arrott plots for (c) $x$ = 0.05 and 0.07 and (d) 0.15, and 0.2 samples at 5 K. The solid lines are the linear extrapolation of the Arrott plots at high fields.

FIG. 4: (Color online) Neutron diffraction patterns for $x$ = 0.03, 0.05, 0.07, and 0.2 samples at below and above the magnetic ordering temperatures. The open circles show the observed patterns. The solid lines represent the Rietveld refined patterns. The difference between observed and calculated patterns is also shown at the bottom of each panel by solid lines. The vertical bars indicate the allowed Bragg peaks position for chemical (top row) and magnetic (bottom row) phase. Symbol (*) marks the additional AFM Bragg peaks.

FIG. 5: (Color online) Neutron diffraction patterns for $x$ = 0.15 sample at various temperatures. The open circles show the observed patterns. The solid lines represent the Rietveld refined patterns. The difference between observed and calculated patterns is also shown at the bottom of each panel by solid lines. The vertical bars indicate the position of allowed Bragg peaks (top for chemical and bottom for magnetic phases). Symbol (*) marks the additional AFM Bragg peaks. (d) Difference pattern obtained by subtraction of the neutron diffraction pattern at 250 K from the neutron diffraction pattern at 1.5 K which shows both AFM and FM contributions. (e) Difference pattern obtained by subtraction of the neutron diffraction pattern at 250 K from the neutron diffraction pattern at 50 K which shows the FM contribution. (f) Neutron diffraction pattern for $x$ = 0.15 sample at 300 K. (g) Variation of intensities with temperature for (111), (200) and (220) Bragg peaks.

FIG. 6: (Color online) Variation of the AFM and the FM phases with change in the Ni concentration at ~ 5 K. Open square and open triangle represent the FM and AFM phase fractions derived from magnetization data, respectively. (+) and (×) represent the FM and AFM phase fractions derived from neutron diffraction data, respectively.

FIG. 7: (Color online) Temperature dependence of the transmitted neutron beam polarization $P$ at an applied field of 50 Oe for $x$ = 0.03, 0.05, 0.07, and 0.15 samples. Inset enlarges the temperature dependence of polarization $P$ for $x$ = 0.03 and 0.05 samples.

FIG. 8: Magnetic phase diagram of the $Cu_{1-x}Ni_xMnSb$ Heusler alloys series from $x$ = 0.03 to 1. Solid circle denotes the Curie temperature for the series taken from Ref. 19. Hollow square and hollow triangle denote the Curie temperature and Néel temperature, respectively, of the samples from the present work.

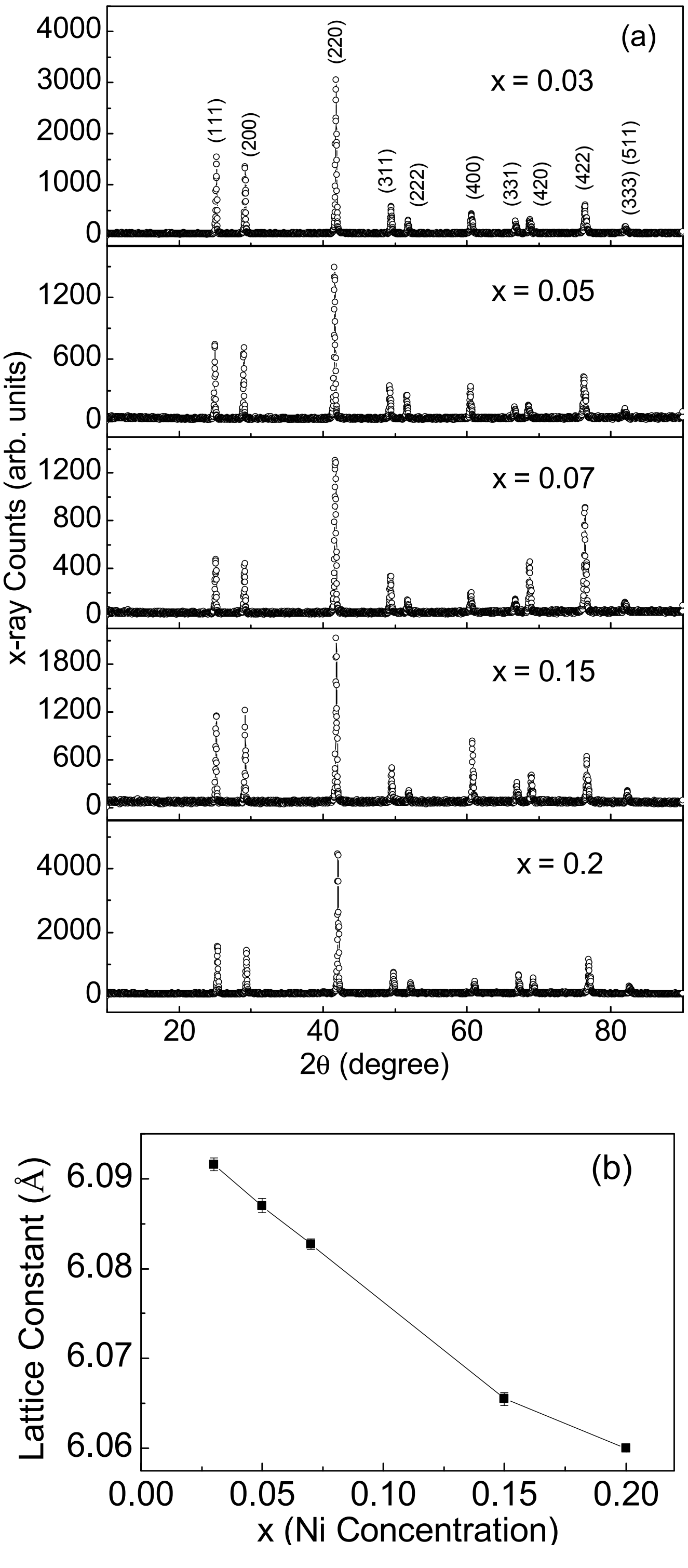

Figure 1
Halder *et al*.

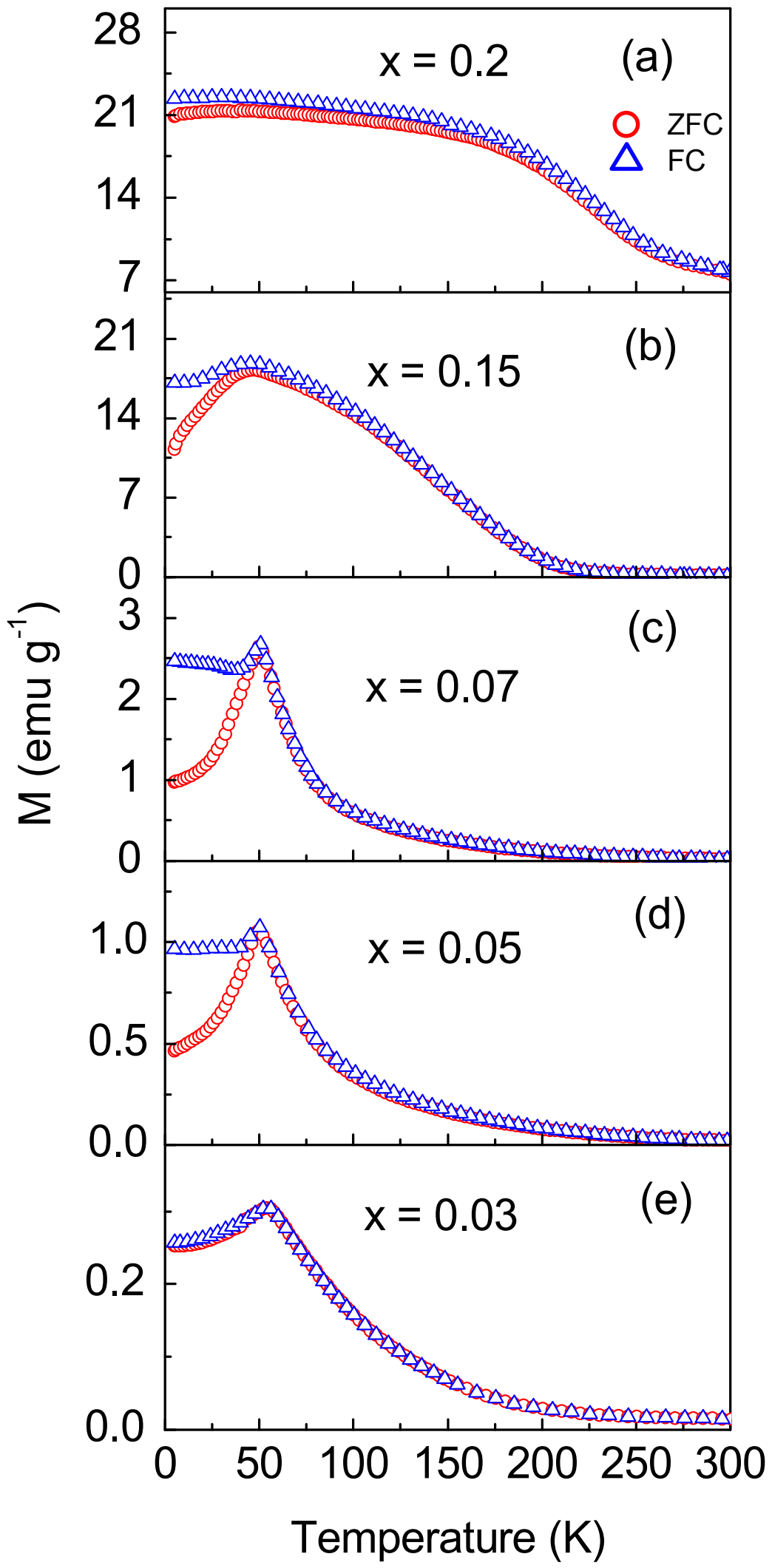

Figure 2
Halder *et al*.

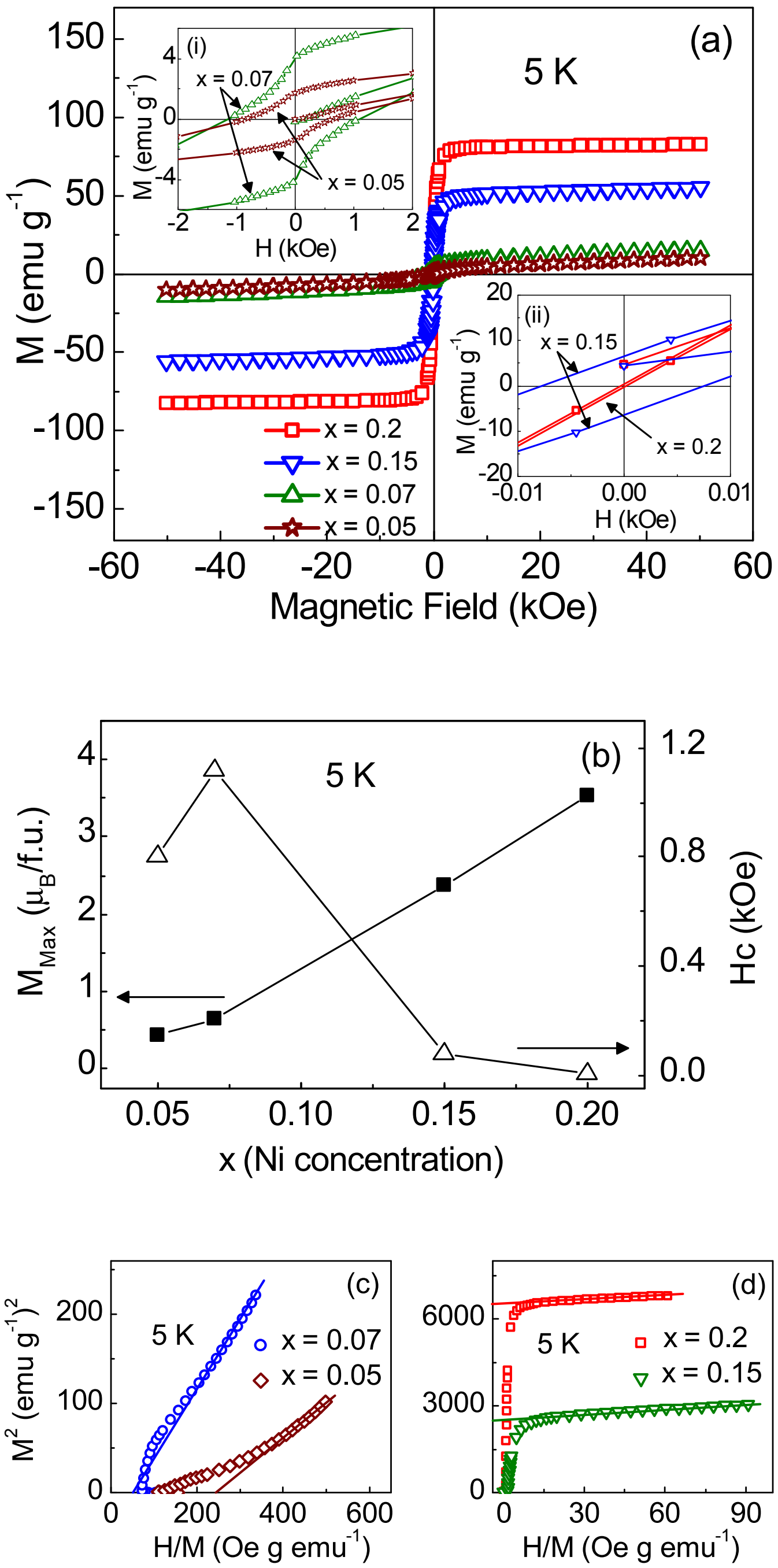

Figure 3
Halder *et al.*

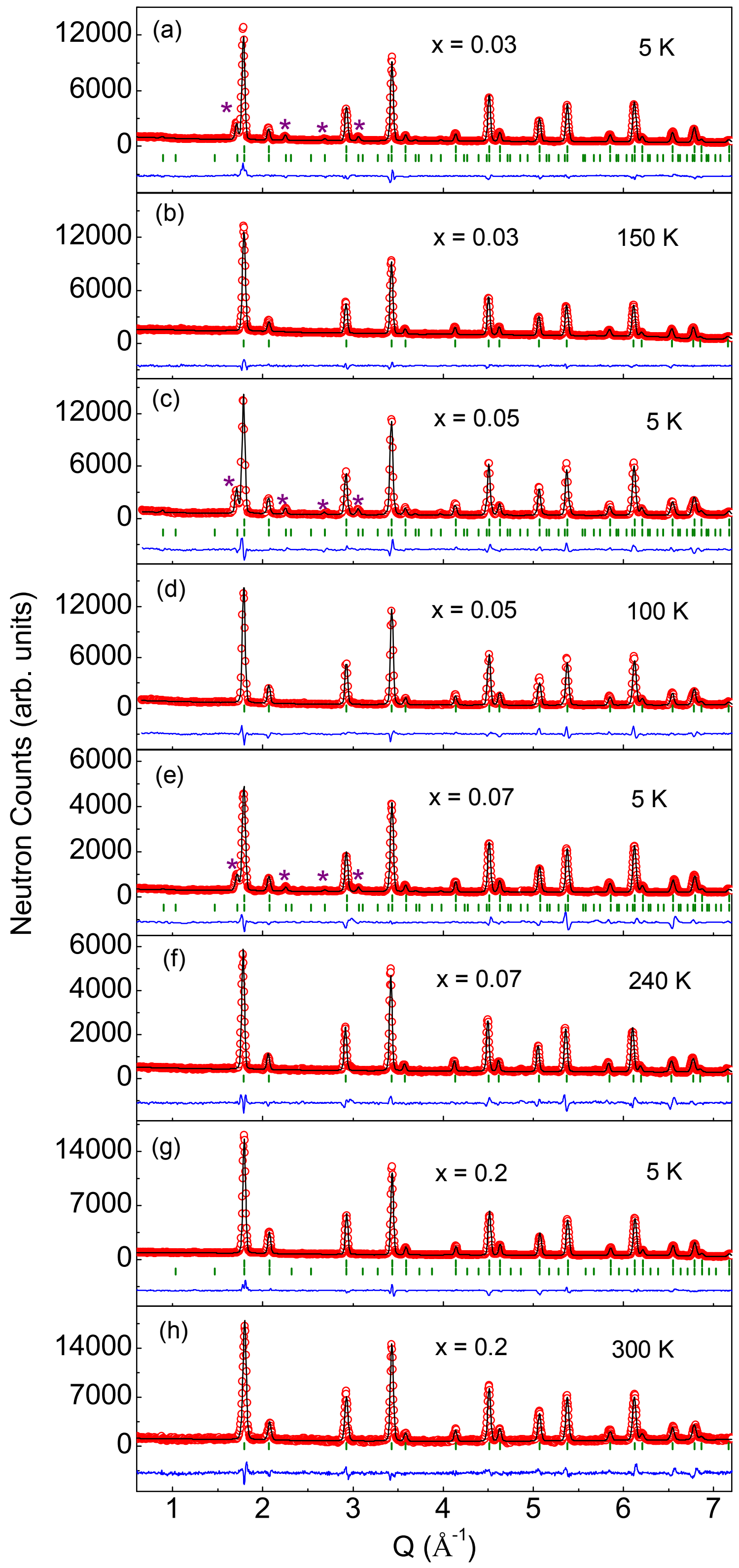

Figure 4
Halder *et al*.

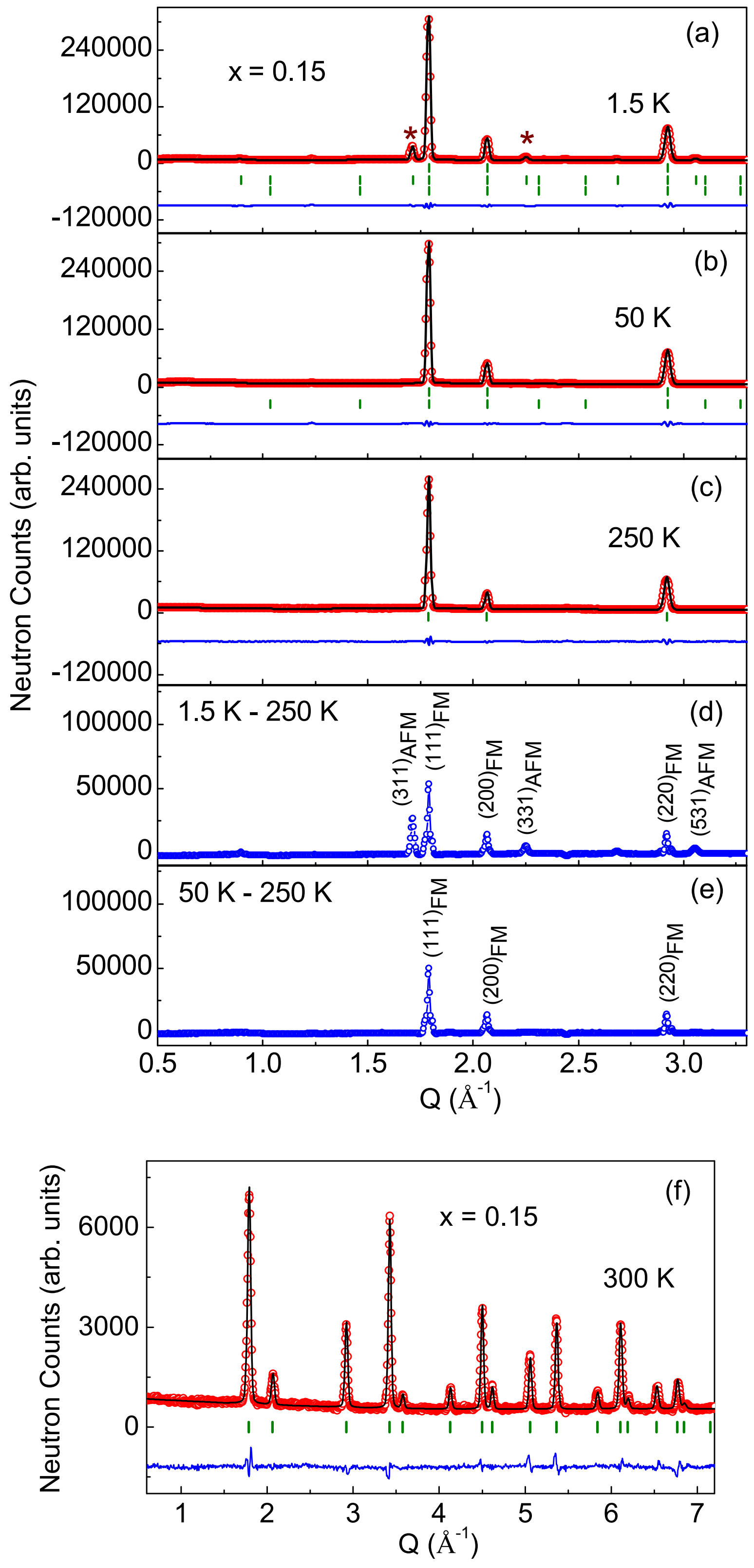

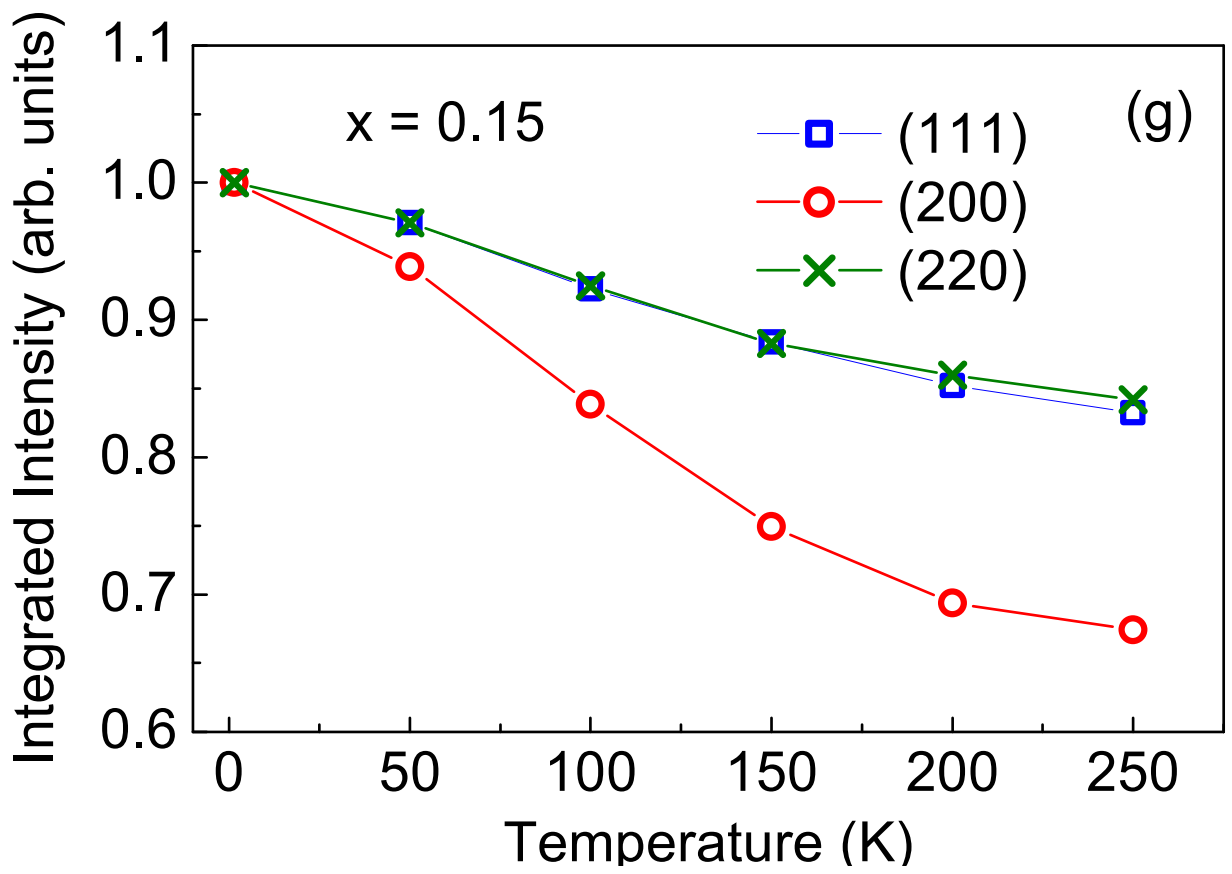

Figure 5
Halder *et al.*

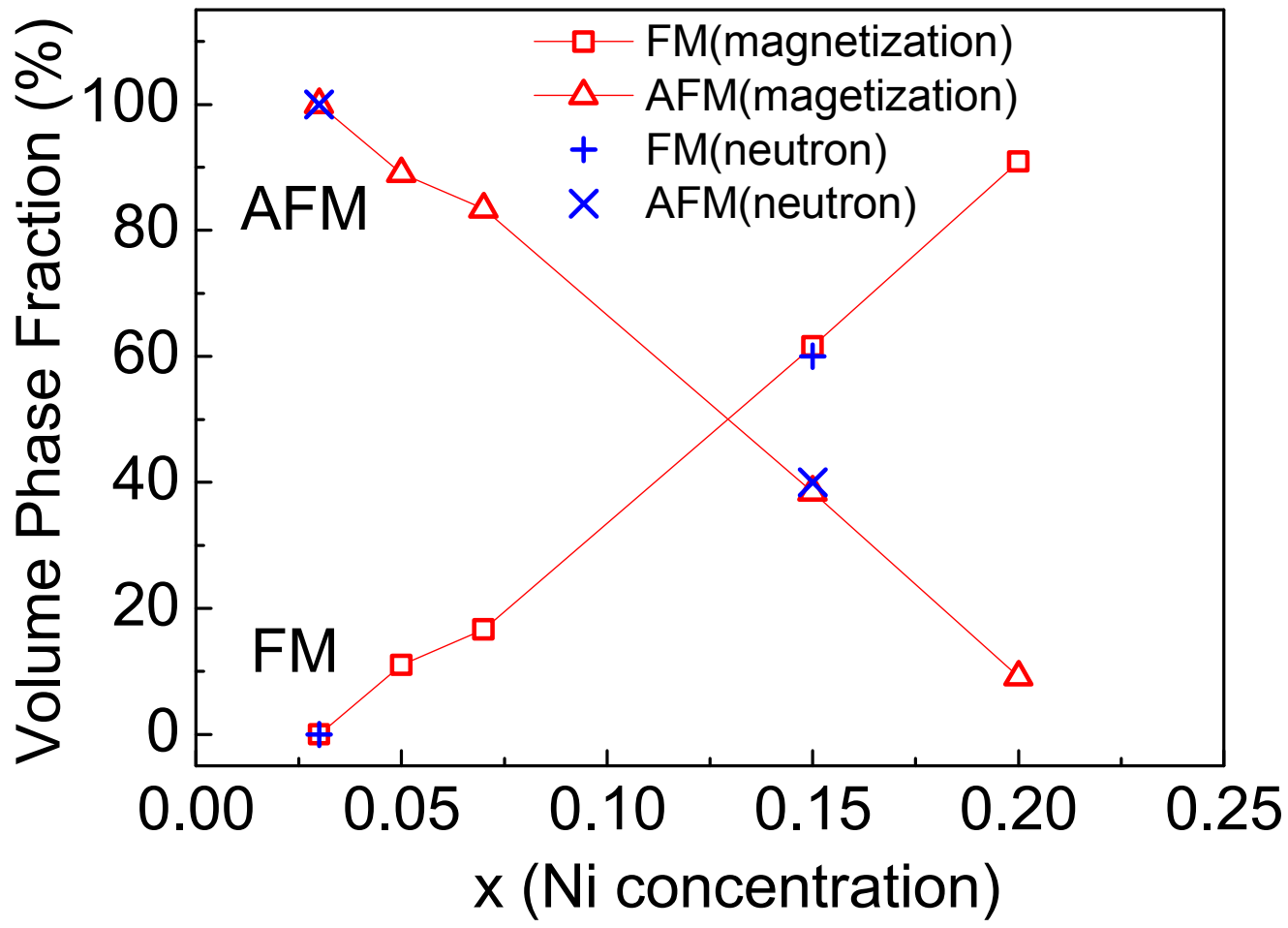

Figure 6
Halder *et al.*

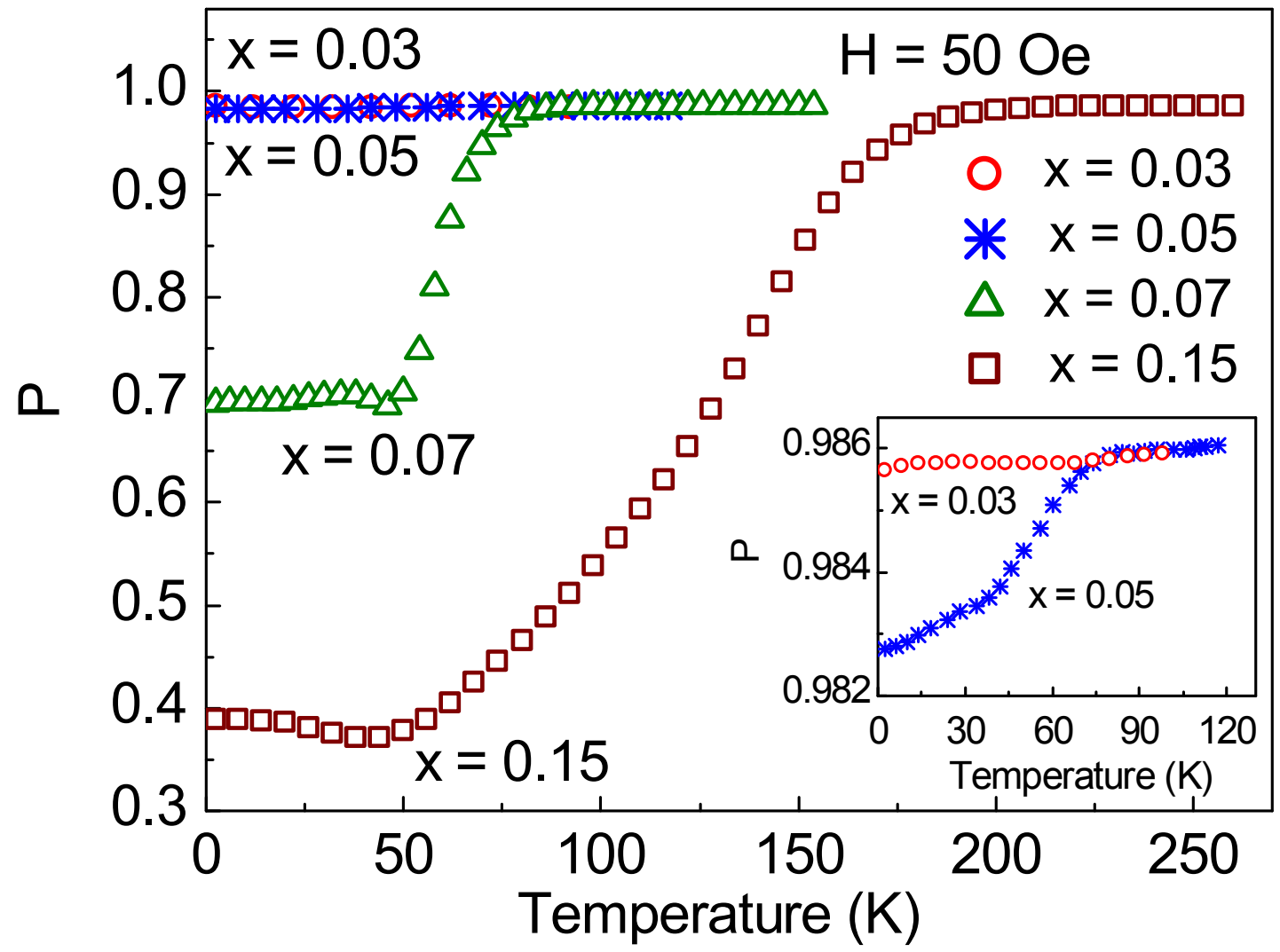

Figure 7
Halder *et al.*

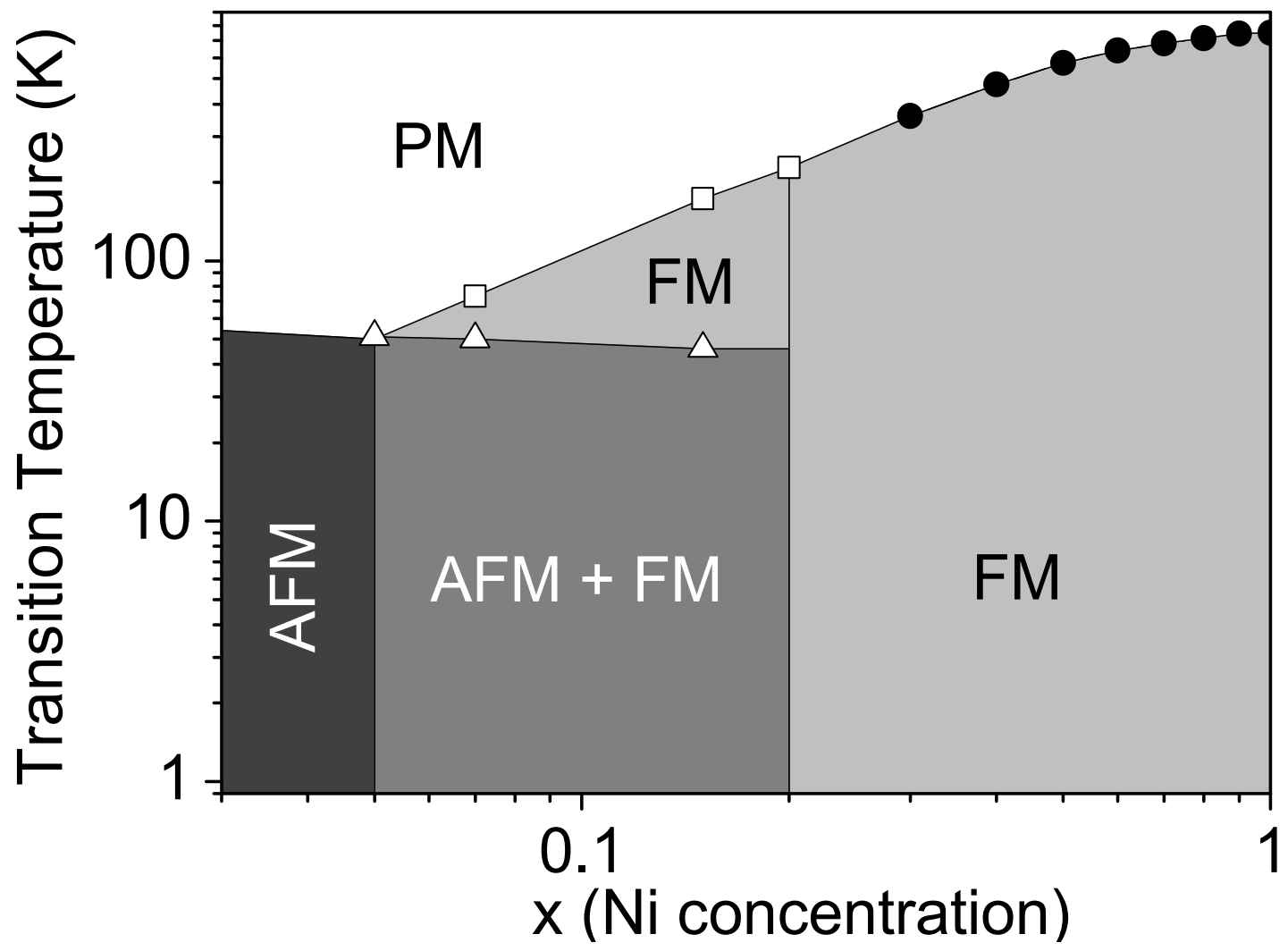

Figure 8
Halder *et al.*